\documentclass[aps, prb, showpacs, twocolumn, amsmath, letterpaper, groupedaddress, superscriptaddress]{revtex4}
\usepackage{graphicx}
\usepackage{amssymb}
\begin{document}
\title{Investigation of superconducting and normal-state properties of the filled-skutterudite system PrPt$_{4}$Ge$_{12-x}$Sb$_{x}$}

\author{I. Jeon}
\affiliation{Materials Science and Engineering Program, University of California, San Diego, La Jolla, California 92093, USA}
\affiliation{Center for Advanced Nanoscience, University of California, San Diego, La Jolla, California 92093, USA}

\author{K. Huang}
\altaffiliation[Present Address: ]{State Key Laboratory of Surface Physics, Department of Physics, Fudan University, Shanghai 200433, China}
\affiliation{Materials Science and Engineering Program, University of California, San Diego, La Jolla, California 92093, USA}
\affiliation{Center for Advanced Nanoscience, University of California, San Diego, La Jolla, California 92093, USA}

\author{D. Yazici}
\altaffiliation[Present Address: ]{Faculty of Health Sciences, Artvin Coruh University, Artvin 08100, Turkey}
\affiliation{Center for Advanced Nanoscience, University of California, San Diego, La Jolla, California 92093, USA}
\affiliation{Department of Physics, University of California, San Diego, La Jolla, California 92093, USA}

\author{N. Kanchanavatee}
\altaffiliation[Present Address: ]{Department of Physics, Faculty of Sciences, Chulalongkorn University, Pathumwan, Bangkok 10330, Thailand}
\affiliation{Center for Advanced Nanoscience, University of California, San Diego, La Jolla, California 92093, USA}
\affiliation{Department of Physics, University of California, San Diego, La Jolla, California 92093, USA}

\author{B. D. White}
\altaffiliation[Present Address: ]{Department of Physics, Central Washington University, 400 E. University Way, Ellensburg, WA 98926-7442, USA}
\affiliation{Center for Advanced Nanoscience, University of California, San Diego, La Jolla, California 92093, USA}
\affiliation{Department of Physics, University of California, San Diego, La Jolla, California 92093, USA}

\author{P.-C. Ho}
\affiliation{Department of Physics, California State University Fresno, Fresno, California 93740, USA}

\author{S. Jang}
\altaffiliation[Present Address: ]{Advanced Light Source, Lawrence Berkeley Laboratory, Berkeley, California 94720, USA}
\affiliation{Materials Science and Engineering Program, University of California, San Diego, La Jolla, California 92093, USA}
\affiliation{Center for Advanced Nanoscience, University of California, San Diego, La Jolla, California 92093, USA}

\author{N. Pouse}
\affiliation{Center for Advanced Nanoscience, University of California, San Diego, La Jolla, California 92093, USA}
\affiliation{Department of Physics, University of California, San Diego, La Jolla, California 92093, USA}

\author{M. B. Maple}
\email[Corresponding Author: ]{mbmaple@ucsd.edu}
\affiliation{Materials Science and Engineering Program, University of California, San Diego, La Jolla, California 92093, USA}
\affiliation{Center for Advanced Nanoscience, University of California, San Diego, La Jolla, California 92093, USA}
\affiliation{Department of Physics, University of California, San Diego, La Jolla, California 92093, USA}



\begin{abstract}
We report a study of the superconducting and normal-state properties of the filled-skutterudite system PrPt$_{4}$Ge$_{12-x}$Sb$_x$. Polycrystalline samples with Sb concentrations up to $x =$ 5 were synthesized and investigated by means of x-ray diffraction, electrical resistivity, magnetic susceptibility, and specific heat measurements. We observed a suppression of superconductivity with increasing Sb substitution up to $x =$ 4, above which, no signature of superconductivity was observed down to 140 mK. The Sommerfeld coefficient, $\gamma$, of superconducting specimens decreases with increasing $x$ up to $x =$ 3, suggesting that superconductivity may depend on the density of electronic states in this system. The specific heat for $x =$ 0.5 exhibits an exponential temperature dependence in the superconducting state, reminiscent of a nodeless superconducting energy gap. We observed evidence for a weak ``rattling'' mode associated with the Pr ions, characterized by an Einstein temperature $\Theta_{\mathrm{E}} \sim$ 60 K for 0 $\leq x \leq$ 5; however, the rattling mode may not play any role in suppressing superconductivity.  
\end{abstract}
\date{\today}


\maketitle


\section{Introduction}
Filled-skutterudite compounds with the generic chemical formula $MT_4 X_{12}$, where $M$ is an alkali metal, alkaline earth, lanthanide, or actinide, $T$ is a transition metal from the Fe or Co column, and $X$ is a pnictogen~[\onlinecite{Jeitschko77}], have been of interest to physicists and chemists worldwide due to the various types of strongly correlated electron behaviors they exhibit and their potential for use in applications; intriguing properties studied in filled skutterudite compounds include conventional BCS-type superconductivity, unconventional superconductivity, Kondo-lattice behavior, valence fluctuations, non-Fermi liquid behavior, heavy-fermion behavior, Kondo-insulator behavior, metal-insulator transitions, magnetic ordering, spin fluctuations, and quadrupolar order~[\onlinecite{Shirotani97, Sales02, Bauer02a, Bauer02b, Maple02, MacLaughlin02, Aoki03, Vollmer03, Suderow04, Maple05, Yuhasz06, Maple07, Maple08, Sato09, Shu09}]. Filled skutterudites also have demonstrated potential viability for use in thermoelectric applications~[\onlinecite{Sales02, Maple08, Sato09}].  

\indent Among the filled-skutterudite compounds, Pr-based systems, in particular, exhibit unusual physical properties~[\onlinecite{Shirotani97, Bauer02a, Sato03, Butch05, Yuhasz06}], including a metal-insulator transition and a low-field ordered phase in PrRu$_4$P$_{12}$ and PrFe$_4$P$_{12}$, respectively~[\onlinecite{Aoki02, Sato03}]. The heavy-fermion superconducting state in PrOs$_4$Sb$_{12}$ has attracted considerable interest and was the first such state to be discovered in a Pr-based system. It has a very large Sommerfeld coefficient, $\gamma \sim$ 500 mJ/mol K$^2$~[\onlinecite{Bauer02a, Maple05}], and exhibits unconventional superconductivity in which there is evidence for time-reversal symmetry breaking~[\onlinecite{Aoki03}], multiple superconducting bands~[\onlinecite{Seyfarth06, Hill08, Shu09}], point nodes in the energy gap~[\onlinecite{Izawa03, Hill08}], and potential spin-triplet pairing of electrons~[\onlinecite{Chia03}]. 

\indent Recently, a new class of filled-skutterudite compounds with the chemical formula \textit{M}Pt$_{4}$Ge$_{12}$ has been synthesized and studied~[\onlinecite{Bauer07, Bauer08, Gumeniuk08, Toda08, Maisuradze09, Kanetake10, Gumeniuk10}]. The compounds with $M =$ Sr, Ba, La, Pr, and Th have all been found to display superconductivity. For (Sr,Ba)Pt$_4$Ge$_{12}$, the superconducting transition temperatures, $T_c$, are $\sim$ 5.4 and 5.1 K, respectively, and BCS-like superconductivity is observed, originating from the Pt-Ge cage structure. For (La,Pr)Pt$_4$Ge$_{12}$, superconductivity occurs at relatively high temperatures of $T_c \sim$ 8.2 and 7.9 K, respectively. Both LaPt$_4$Ge$_{12}$ and ThPt$_4$Ge$_{12}$ exhibit conventional BCS-type superconductivity; however, ThPt$_4$Ge$_{12}$ was also found to be a clean-limit strong-coupling superconductor with $T_c =$ 4.6 K.

\indent In contrast, the compound PrPt$_4$Ge$_{12}$ displays unconventional superconductivity that is similar to that of PrOs$_4$Sb$_{12}$ in several ways. Transverse muon spin relaxation ($\mu$SR) and specific heat measurements suggest point nodes in the superconducting energy gap of PrPt$_4$Ge$_{12}$, and zero-field $\mu$SR measurements provide evidence for time-reversal symmetry breaking in the superconducting state~[\onlinecite{Maisuradze09, Maisuradze10}]. The compound PrPt$_4$Ge$_{12}$ exhibits a similar type of multiband unconventional superconductivity~[\onlinecite{Chandra12, Nakamura12, Zhang13}]. A recent study of Ce substitution into the filler sites for Pr ions in PrPt$_4$Ge$_{12}$ reported the suppression of superconductivity with increasing Ce concentration and suggested a crossover from a nodal to nodeless superconducting energy gap or the suppression of multiple superconducting energy bands, revealing a single, robust BCS-type superconducting energy gap~[\onlinecite{Huang14}]. 

\indent A few studies attempting to understand the effect of chemical substitution within the Pt-Ge cage have been conducted. Upon substitution of Au for Pt in BaPt$_4$Ge$_{12}$, the electronic density of states and superconducting transition temperature $T_c$ increase with increasing Au concentration~[\onlinecite{Gumeniuk08a}]. The effect of Sb substitution for Ge has also been studied for several Pt-Ge based skutterudite compounds: The compound CePt$_4$Ge$_{12}$ is close to a boundary between Ce intermediate valence and Kondo-lattice behavior~[\onlinecite{Toda08, Gumeniuk11}]. By substituting Sb for Ge, CePt$_4$Ge$_{12}$ is tuned from a nearly intermediate-valent paramagnet, through a non-Fermi liquid phase, and into an antiferromagnetically-ordered phase with localized Ce 4\textit{f} magnetic moments~[\onlinecite{Nicklas12, White14}]. A rapid suppression of superconductivity in the system LaPt$_4$Ge$_{12-x}$Sb$_x$ with increasing $x$ is observed, accompanied by a decrease of charge-carrier density and an increase of the Seebeck effect at room temperature by about one order of magnitude~[\onlinecite{Humer13}].

\indent Motivated by previous studies, and in order to further investigate the unresolved nature of superconductivity in the compound PrPt$_4$Ge$_{12}$, we undertook a systematic study of the PrPt$_4$Ge$_{12-x}$Sb$_{x}$ system. The evolution of superconducting properties with increasing Sb concentration was investigated by means of x-ray diffraction, electrical resistivity in zero and applied magnetic fields, DC and AC magnetic susceptibility, and specific heat measurements. We observed a suppression of superconductivity with positive curvature in a plot of $T_c$ versus Sb concentration; above $x =$ 4, there is no evidence for superconductivity down to 140 mK. Our results from specific heat measurements is similar to that of previous studies, suggesting a possible crossover from a nodal to a nodeless superconducting energy gap or from multiple energy gaps to a single BCS-type superconducting energy gap~[\onlinecite{Chandra12,Nakamura12,Zhang13,Huang14}]. Evidence for a ``rattling'' mode for the Pr ions was observed throughout the series, but does not seem to have a strong effect on superconductivity. Conjectures about an observed feature or phase transition of unknown origin above $x =$ 4 are also discussed.  

\section{Experimental Details}
Polycrystalline specimens of PrPt$_4$Ge$_{12-x}$Sb$_x$ were synthesized by arc-melting on a water-cooled copper hearth under an Ar atmosphere with a Zr getter. The starting materials were obtained from Pr ingots (Alfa Aesar 99.9\%), Pt sponge (99.9999+\%), Ge pieces (Alfa Aesar 99.9999+\%), and Sb pieces (Alfa Aesar 99\%). These starting materials were weighed out in accordance with the stoichiometric ratios and arc-melted, turned over, and arc-melted again a total of five times to promote homogeneity of the samples. The arc-melted boules were then annealed in sealed quartz ampoules (containing 150 torr of Ar at room temperature) for 336 hours at 750 $^{\circ}$C. The crystal structure was determined by x-ray powder diffraction (XRD) using a Bruker D8 Discover x-ray diffractometer with Cu-K$_{\alpha}$ radiation, and XRD patterns were analyzed via Rietveld refinement using the GSAS+EXPGUI software package~[\onlinecite{GSAS1,GSAS2}]. The electrical resistivity was measured from 1.1 K to 300 K using a standard four-wire method with a Linear Research LR700 AC resistance bridge in a home-built probe in a liquid $^4$He Dewar, and down to 140 mK using a commercial $^3$He-$^4$He dilution refrigerator. Magnetic susceptibility measurements were performed between 2 K and 300 K in magnetic fields up to 7 T using a Quantum Design Magnetic Property Measurement System (MPMS). Alternating current magnetic susceptibility measurements were made down to $\sim$ 1.1 K in a liquid $^4$He Dewar using home-built mutual inductance coils. Specific heat measurements were carried out at temperatures down to 1.8 K with a Quantum Design Physical Property Measurement System (PPMS) DynaCool.

\section{Results}

\subsection{X-ray diffraction}

\begin{figure}
\includegraphics[width=8.5cm]{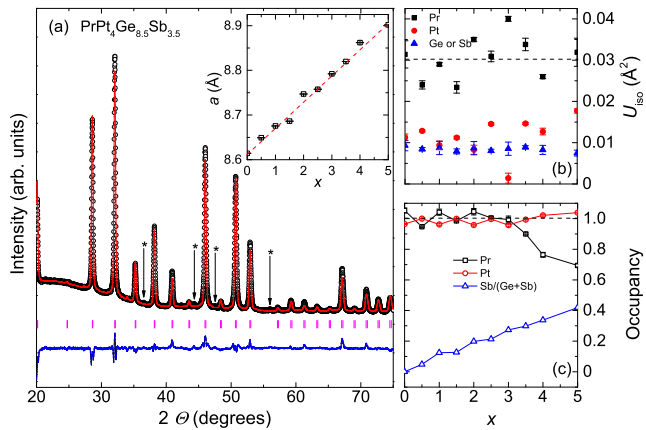}
\caption{(Color online) (a) X-ray diffraction pattern for PrPt$_4$Ge$_{8.5}$Sb$_{3.5}$. The black circles represent the experimental data and the red line represents the fit from the Rietveld refinement of the data. The purple vertical marks indicate the position of expected Bragg reflections for the refined PrPt$_4$Ge$_{8.5}$Sb$_{3.5}$ crystal structure and the blue line at the bottom is the difference between observed and calculated intensities. The stars indicate Bragg reflections associated with a Pr or PtSb$_{2}$ impurity phase. The inset shows a plot of the lattice parameter $a$ versus nominal antimony concentration \textit{x}. The red dashed line is a guide to the eye. (b) Thermal displacement parameter, $U_{\mathrm{iso}}$, of elements versus Sb concentration. Pr atoms have large values compared with Pt and Ge/Sb atoms. (c) Occupancies of Pr and Pt crystallographic positions and the ratio of Sb to Ge + Sb versus Sb concentration $x$. A significant decrease of Pr site occupancy is observed for $x \geq$ 3.5.}
\label{fig:XRD}
\end{figure}

Figure 1 shows results from XRD data for PrPt$_4$Ge$_{12-x}$Sb$_x$ (0 $\leq x \leq$ 5). All of the XRD patterns are well indexed with the cubic filled-skutterudite crystal structure with space group $Im \bar 3$. The conventional residual parameters, $R_{\mathrm{p}}$, are in the range from 0.0910 to 0.1595 for $x \leq$ 5. Figure~\ref{fig:XRD}(a) shows a representative XRD pattern for the PrPt$_4$Ge$_{12-x}$Sb$_x$ system (for $x =$ 3.5) and the resultant fit from the Rietveld refinement. The stars indicate the presence of Pr and/or PtSb$_2$ impurity phases, the amount of which gradually increases with $x$ from $\sim$ 0.2 to 1\% by mass ratios for 3 $\leq x \leq$ 5. Additionally, Ge, PrSb$_2$, and PrPtGe impurities were observed; however, the amount of these impurities is less than 1\% by mass ratio, suggesting very weak effects on the properties of these samples. For samples with $x \geq$ 6, the dominant phase is PtSb$_2$; the mass ratios of PtSb$_2$ by Rietveld refinement are 6\% and 52\% for $x =$ 6 and 7, respectively, indicating the solubility limit is near or just beyond $x =$ 6. This result is consistent with the observation that the lattice parameter $a$ exhibits a plateau for $x \geq$ 5 (data not shown). Since the atomic radius of Sb is larger than that of Ge, it is expected that $a$ increases with increasing Sb concentrations. As seen in Figure~\ref{fig:XRD}(b), we observed that the thermal displacement parameters, $U_{\mathrm{iso}}$, for Pr atoms are large compared with values of Pt or Ge/Sb atoms, which is a common feature in filled-skutterudite systems. Figure~\ref{fig:XRD}(c) displays the occupancies of the Pr and Pt crystallographic positions and the relative ratios of Sb to Sb+Ge occupancies, suggesting that the Pr sites are not fully occupied down to $\sim$ 0.7 for $x \geq$ 3.5. A similar result is occasionally observed in other filled skutterudite compounds~[\onlinecite{Bauer02, Bauer04}]. This result is also consistent with the increase of Pr-based impurity phases for samples with $x \geq$ 3.

\subsection{Electrical Resistivity}
\begin{figure}
\includegraphics[width=8.5cm]{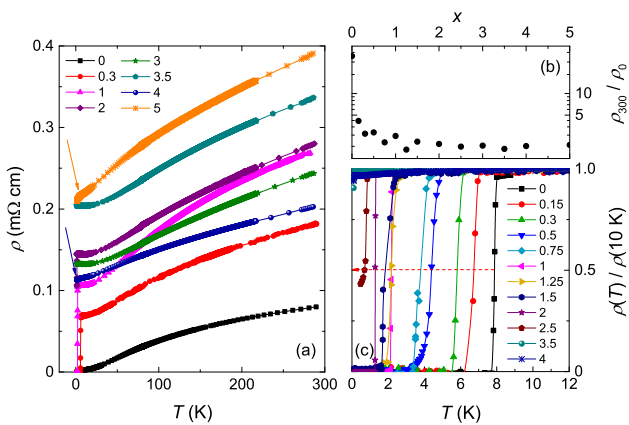}
\caption{(Color online) (a) Electrical resistivity data $\rho$ versus $T$ for selected PrPt$_4$Ge$_{12-x}$Sb$_x$ samples with Sb concentrations $x$ shown in the legend. The arrows indicate a weak downturn feature for $x =$ 4 and 5, possibly due to an impurity phase. (b) A semilogarithmic plot of the residual resistivity ratio RRR ($\rho _{300} / \rho _{0}$) versus $x$ for PrPt$_4$Ge$_{12-x}$Sb$_x$. The decrease in RRR with increasing $x$ displays a positive curvature. (c) $\rho (T)$, normalized to its value at 10 K, versus $x$ for selected PrPt$_4$Ge$_{12-x}$Sb$_x$ samples. Superconducting transition temperatures $T_c$ decrease with increasing $x$ and onsets of superconductivity are observed for $x =$ 3.5 and 4. The red dashed arrow is a guide to the eye.}
\label{fig:RHO}
\end{figure}

Electrical resistivity, $\rho$($T$), data taken in zero magnetic field are shown in Fig.~\ref{fig:RHO}. All concentrations of Sb display metallic behavior, as shown in Fig.~\ref{fig:RHO}(a). We show only representative concentrations for visual clarity. The residual resistivity ratio, RRR, versus $x$ is presented in Fig.~\ref{fig:RHO}(b); a semilogarithmic plot of ($\rho _{300}$/$\rho _{0}$) versus $x$, where $\rho _{300}$ is the room temperature resistivity and $\rho _{0}$ is the resistivity value right above the superconducting transitions, shows a rapid decrease with increasing $x$ for 0 $\leq x \leq$ 1, consistent with increased disorder produced by the substitution of Sb for Ge in PrPt$_4$Ge$_{12}$. A general trend of increasing residual resistivity $\rho_0$ is observed with increasing Sb concentration; however, there is some scatter due to uncertainties in the measurement of the geometrical factors of the resistivity samples. Figure~\ref{fig:RHO}(c) displays $\rho(T)$ normalized to its value at 10 K versus $x$. $T_c$ was defined as the temperature where $\rho$($T$)/$\rho_{10}$ drops to 50\% of its value (i.e., 0.5), and the width of the transition was characterized by the temperatures where $\rho$($T$)/$\rho_{10}$ is 0.9 and 0.1. Even though the superconducting transitions are slightly broadened for $x =$ 0.5, 1.5, and 2.5, all transitions are still relatively sharp, indicating good sample homogeneity or small amounts of impurity phases in the samples. $T_c$ is suppressed more rapidly with initial Sb substitution with the effect becoming weaker with increasing $x$, similar to the behavior of the RRR versus $x$. Superconductivity onsets are observed for $x =$ 3.5 and 4 in measurements performed down to 140 mK. 

\begin{figure}
\includegraphics[width=8.5cm]{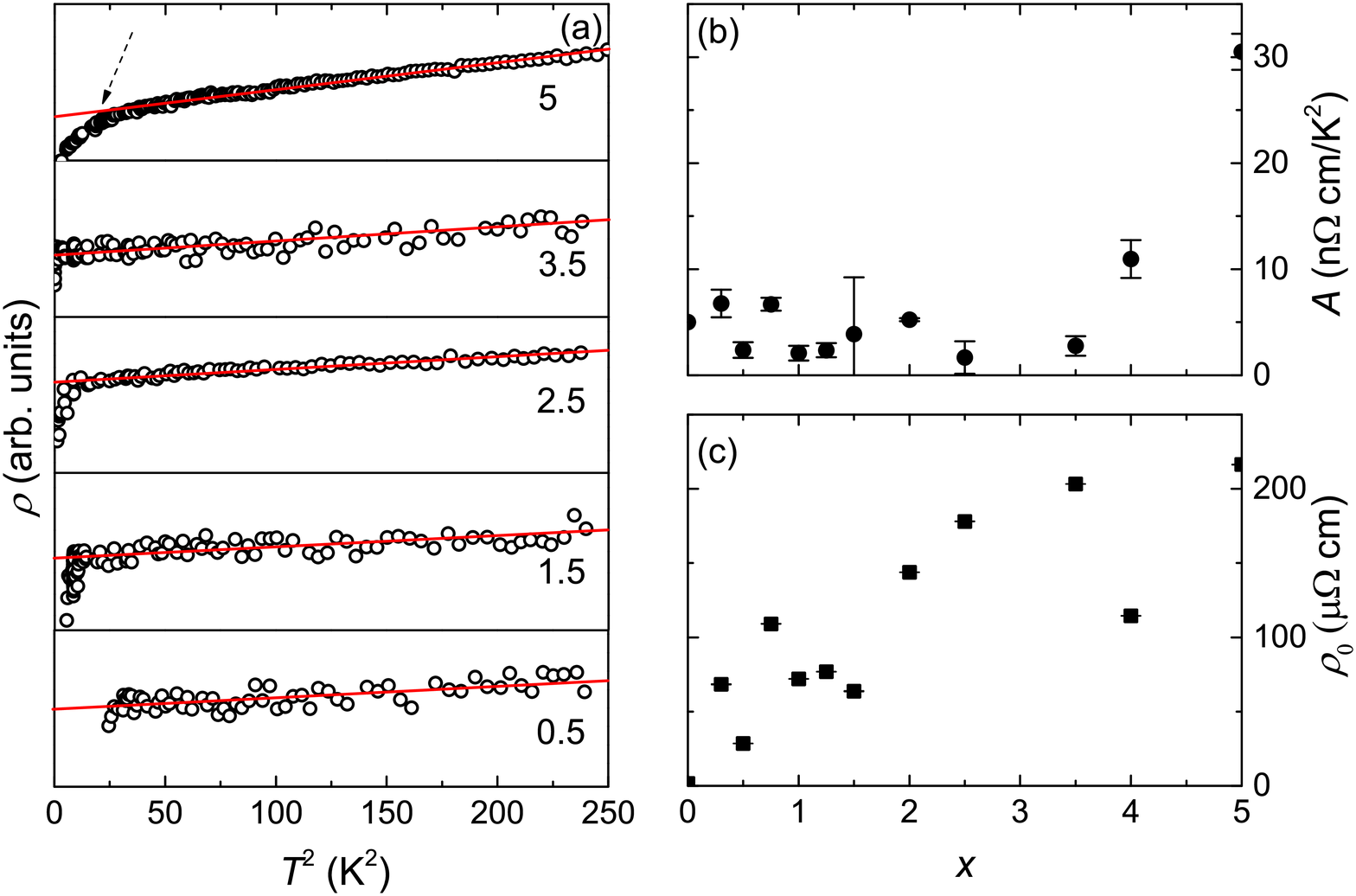}
\caption{(Color online) (a) A plot of $\rho$ versus $T^2$ for selected concentrations with different y-scales for visual clarity. Power-law fits of $\rho (T) = \rho _0 + AT^2$ were performed up to 250 K$^2$, indicated by the red solid lines. The dashed arrow indicates the deviation from the fit, possibly due to an impurity phase. (b) Coefficient $A$ versus $x$. $A$ scatters between 2 and 7 n$\Omega$ cm/K$^2$ and increases rapidly for $x \geq$ 4. (c) The residual resistivity, $\rho_0$, from the fits versus $x$ is shown. $\rho_0$ increases rapidly from $\sim$2 to $\sim$216 $\mu \Omega$ cm, consistent with disorder introduced by Sb substitution into PrPt$_4$Ge$_{12}$.}
\label{fig:nT2}
\end{figure}

Figure~\ref{fig:nT2}(a) shows $\rho$ versus $T^2$ for representative concentrations, with different y-scales for visual clarity. The red solid lines represent least squares fits to the data with the formula $\rho (T) =\rho_0 + AT^2$ in the temperature range between $T_c ^2$ and 250 K$^2$, suggesting Fermi-liquid behavior in PrPt$_4$Ge$_{12-x}$Sb$_x$ for 0 $\leq x \leq$ 5~[\onlinecite{Huang14}]. Deviations from the fits were observed for the $x =$ 4 and 5 samples, which is probably due to the PtSb$_2$ impurity phase. We also performed a resistivity measurement on the samples with $x =$ 6, 7, and 12 (data not shown); these inhomogeneous samples exhibit the same feature marked by arrows in Figs.~\ref{fig:RHO} and \ref{fig:nT2} at the same temperature and with the same character. For $x =$ 6 and 7, the amount of the PtSb$_2$ phase is large; moreover, the $x= 12$ sample turned out to be the compound PtSb$_2$, with a very small amount of PrPt$_4$Ge$_7$Sb$_5$. The coefficient, $A$, scatters between 2 to 7 n$\Omega$ cm/K$^2$, increasing to 30 n$\Omega$ cm/K$^2$ for $x \geq$ 4. Figure~\ref{fig:nT2}(c) shows that values of the residual resistivity $\rho_0$ versus $x$, obtained from the fits, increase from $\sim$ 2 to 216 $\mu \Omega$ cm; though, the values of $\rho_0$ fluctuate strongly with increasing $x$ due to uncertainties in the measurement of the geometrical factors of the resistivity samples, as we mentioned previously to explain the behavior seen in Fig.~\ref{fig:RHO}(a).

\begin{figure}
\includegraphics[width=8.5cm]{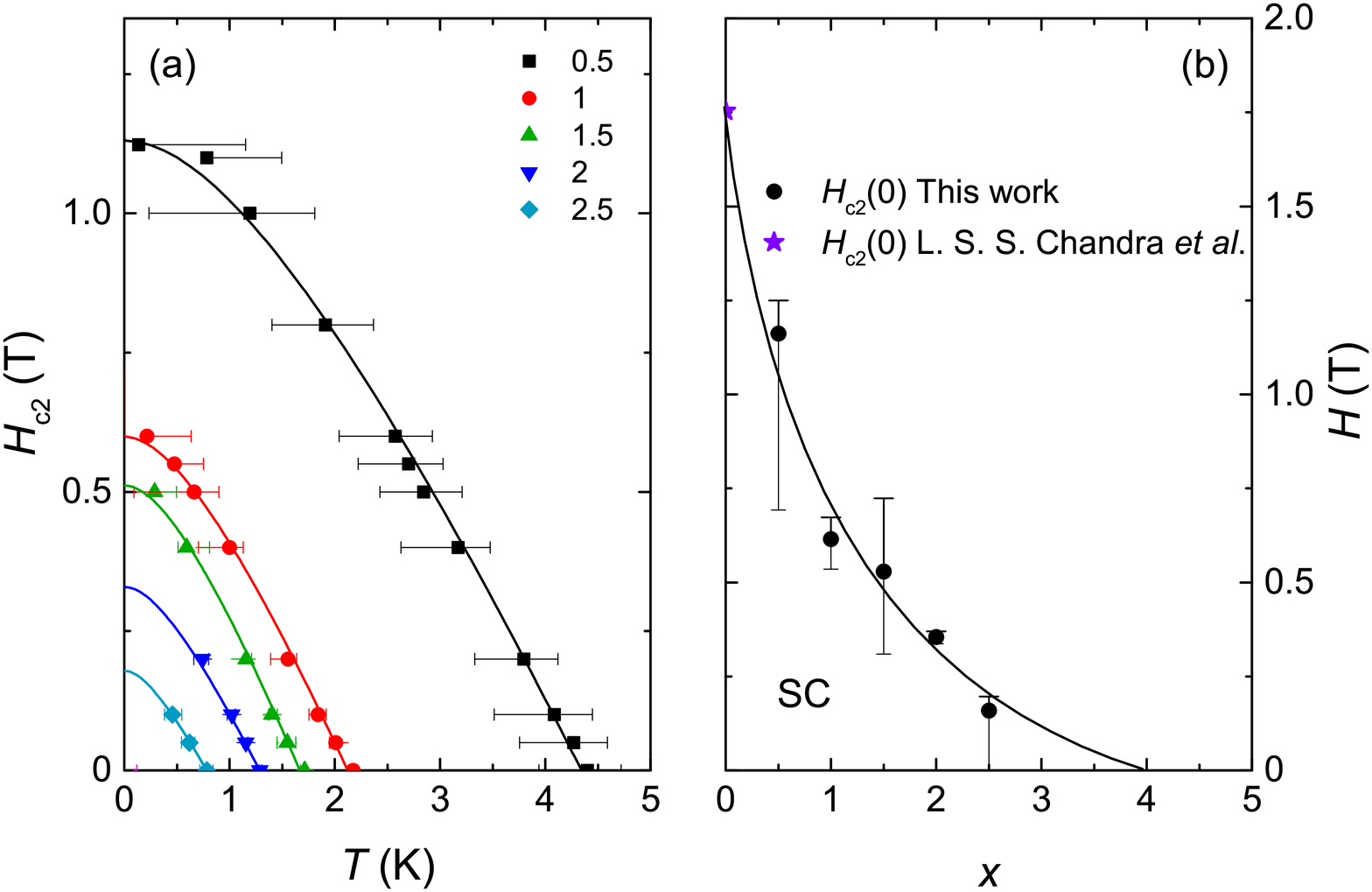}
\caption{(Color online) (a) Temperature-dependence of the upper critical field $H_{\mathrm{c}2}$ for selected samples. The horizontal bars are the transition widths, determined as described in the text. Solid lines are fits of the WHH theory to the data. (b) Magnetic field $H$ - $x$ phase diagram of PrPt$_4$Ge$_{12-x}$Sb$_x$ at $T =$ 0 K. The solid line, which is a guide to the eye, represents the rapid suppression of the 0 K value of the upper critical field, $H_{\mathrm{c}2}$(0), with increasing $x$. The values of $H_{\mathrm{c}2}$(0) were determined from the WHH fits in Fig.~\ref{fig:Hc2}(a).}
\label{fig:Hc2}
\end{figure}

We also measured the upper critical fields, $H_{\mathrm{c}2}(T)$, for selected samples. In Fig.~\ref{fig:Hc2}(a), the data points were determined from the temperature where $\rho(T)$ decreases to 50\% of its value in the normal state just above $T_c$ at fixed magnetic fields, and the width of the transitions was defined using the 10\% and 90\% values of the drop in $\rho (T)$. In general, the superconducting transitions become broader at higher applied magnetic fields. The derivative $(\mathrm{d} H_{\mathrm{c}2} / \mathrm{d}T)_{T=T_c}$ was obtained by fitting straight lines to the data near $H_{\mathrm{c}2} =$ 0 and estimated values are $\sim -$0.39 T/K for all concentrations. Using the Werthamer-Helfand-Hohenberg (WHH) model~[\onlinecite{Werthamer66,Chandra12}], where $H_{\mathrm{c}2} (0)$ is $-0.693 T_c(\mathrm{d}H_{\mathrm{c}2}/T)_{T=T_c}$, the temperature dependence of $H_{\mathrm{c}2}$ was extracted, as shown in Fig. 4(a). Figure~\ref{fig:Hc2}(b) reveals a rapid decrease of $H_{\mathrm{c}2}$(0) with increasing $x$, showing similar behavior to those observed in the rapid suppressions of $T_c$ and the RRR with increasing $x$. It is noteworthy that the curvature of $H_{\mathrm{c}2} (T)$ seems to decrease with increasing $x$, similar to the behavior previously observed in the Pr(Os$_{1-x}$Ru$_x$)$_{4}$Sb$_{12}$ system~[\onlinecite{Ho08}].

\subsection{Magnetic Susceptibility}
\begin{figure}
\includegraphics[width=8.5cm]{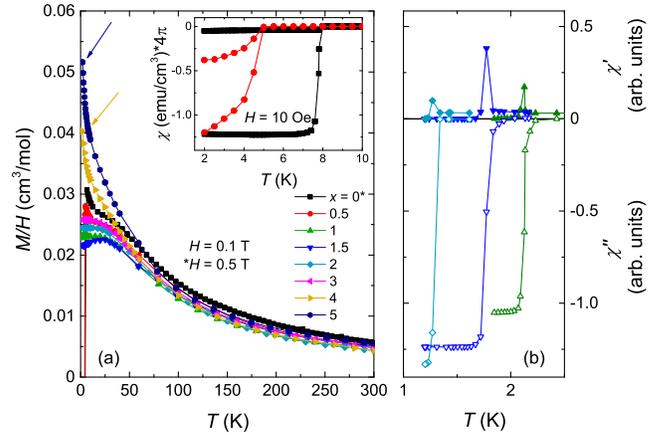}
\caption{(Color online) (a) Magnetization divided by applied magnetic field, $M/H$, versus temperature \textit{T} for PrPt$_4$Ge$_{12-x}$Sb$_x$ with selected concentrations, measured in applied magnetic fields of $H =$ 0.5 T for the sample with $x =$ 0 and $H =$ 1 T for the samples with $x$ $\textgreater$ 0. Solid arrows indicate upturns due to small amounts of paramagnetic impurities, as discussed in the text. The inset highlights the Meissner effect and diamagnetic shielding plotting 4$\pi \chi$ versus $T$ for the $x =$ 0 and 0.5 samples. (b) Alternating current magnetic susceptibility $\chi'$ and $\chi''$ versus temperature $T$ for $x =$ 1, 1.5, and 2. (a), (b) Superconducting volume fractions are close to 1. The deviations from unity are presumably due to not accounting for the demagnetization factor.}
\label{fig:M_H}
\end{figure}

Magnetization divided by applied magnetic field, $M/H$, versus $T$ data are displayed in Fig.~\ref{fig:M_H}(a). We performed measurements under applied magnetic field of $H =$ 0.5 T for the $x =$ 0 sample and 1 T for the rest of the samples containing Sb. The magnitude of $M/H$ decreases with increasing $x$; however, an increase in magnitude and low-temperature upturns were observed for $x \geq$ 4, which are expected from the XRD results. The decrease of the estimated Pr occupancy (see Fig.~\ref{fig:XRD}(c)) and presence of $\sim$1 atomic percent of Pr and PtSb$_2$ impurity phases in $x =$ 4 and 5 samples might be the reasons. 
The inset of Fig.~\ref{fig:M_H}(a) shows superconducting transitions for $x =$ 0 and 0.5 under an applied magnetic field $H =$ 10 Oe. $T_c$ was defined by the point where zero-field-cooled (ZFC) and field-cooled (FC) data deviated from one another. The superconducting volume fractions were estimated from the ZFC $\chi(T)$ data by using the relation 4$\pi \chi \times d$, where $d$ is the molar density of the samples in units of mol/cm$^3$. The superconducting samples with $x =$ 0, 0.5, 1, 1.5, and 2 have volume fractions slightly greater than 1, which results from exclusion of demagnetization factor corrections in this analysis; nevertheless, these results offer strong support for bulk superconductivity in the system PrPt$_4$Ge$_{12-x}$Sb$_x$.

\begin{figure}
\includegraphics[width=8cm]{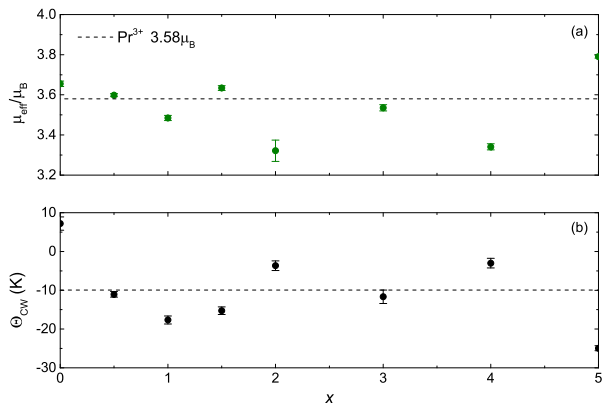}
\caption{(Color online) (a), (b) Effective magnetic moment ($\mu _{eff}/\mu _{B}$) and Curie-Weiss temperature $\Theta_{\mathrm{CW}}$ versus $x$ from the Curie-Weiss fits to $M/H$ data. $\mu _{eff}/\mu _{B}$ values scatter around $\mu_{eff} =$ 3.58$\mu_{B}$, the value expected for the Pr$^{3+}$ free ion, indicated by the dashed line. Both $\mu _{eff}/\mu _{B}$ and $\Theta_{\mathrm{CW}}$ are nearly $x$ independent. The dashed line in (b) is a guide to the eye.}
\label{fig:mu_theta}
\end{figure}

\indent We fit the $M/H$ versus $T$ data in Fig.~\ref{fig:M_H}(a) to a Curie-Weiss law in the temperature range from 75 to 300 K,
\begin{equation}
M/H = C_0/(T - \Theta_{\mathrm{CW}}),
\label{eq:CW}
\end{equation}
where $C_0$ is the Curie constant and $\Theta_{\textrm{CW}}$ is the Curie-Weiss temperature. The average effective magnetic moment, $\mu_{eff}$, of the Pr ion is extracted from the relation $C_0 = \mu^2 _{eff}N_A /$3$k_B$, where $N_A$ is Avogadro's number and $k_B$ is Boltzmann's constant. The best fit values are shown in Fig.~\ref{fig:mu_theta}. Values of $\mu _{eff}$ scatter around 3.58$\mu_{B}$, the value for Pr$^{3+}$ free ions, calculated using Hund's rules. The Curie-Weiss temperatures, $\Theta_{\mathrm{CW}}$, are nearly independent of $x$ with a value of $\sim -$10 K. Real and imaginary components of the alternating current magnetic susceptibility, $\chi'$ and $\chi''$, respectively, for the samples with $x =$ 1, 1.5, and 2 are shown in Fig.~\ref{fig:M_H}(b). Clear signatures of superconductivity were observed and $T_c$ was defined as the temperatures where $\chi''$ drops sharply.

\indent Isothermal magnetization measurements were also performed at 2 K under applied magnetic fields up to 7 T (data not shown). We observed a rapid suppression of superconductivity below 1 T and above that, paramagnetism was observed; however, slight curvatures in the isothermal magnetization for the samples with $x =$ 4 and 5 were seen, possibly due to small concentrations of paramagnetic impurities, as we discussed in previous sections. In order to roughly estimate the concentrations of paramagnetic impurities, Gd was used as a standard impurity and assumed to be located at the Pr sites. This choice was arbitrary; it could be other lanthanide ions on the Pr site or 3$d$ transition metal ions such as Fe on the Pt site. The impurity concentration $(N/V)$ was determined in two ways: Curie law fits to the upturns at low temperatures and the procedure described in reference [\onlinecite{Lukefahr95}], using isothermal magnetization data. For the Curie law fits, we used the relation $N/V = 3C_{\mathrm{0}}k_{\mathrm{B}}/(N_{\mathrm{A}}\mu^2_{\mathrm{eff}})$, where $\mu^2_{\mathrm{eff}}$ is the effective magnetic moment of Gd ions (7.94 $\mu_{\mathrm{B}}$). Both methods yielded estimates of the impurity concentration of about 1 $\sim$ 2\% of the lanthanide ions, consistent with the results from XRD measurements. Since we have evidence for larger amounts of impurities in the samples with high Sb concentrations, we excluded other possibilities for the upturn at low temperatures, such as a change in the crystalline electric field ground state of the Pr ions.

\subsection{Specific Heat}
\begin{figure}
\includegraphics[width=8.5cm]{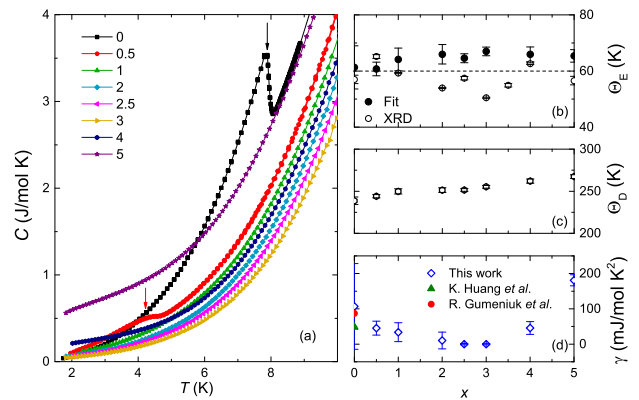}
\caption{(Color online) (a) Specific heat $C$ versus temperature \textit{T} for PrPt$_4$Ge$_{12-x}$Sb$_x$. Arrows indicate superconducting transitions for the $x =$ 0 and 0.5 samples. (b) Einstein temperature, $\Theta_{\mathrm{E}}$, versus $x$. Solid data points are from fits of the specific heat data that include the Einstein model. Open circles are extracted from the relation between $\Theta_{\mathrm{E}}$ and thermal displacement parameter, $U_{\mathrm{iso}}$, from Rietveld refinement of the XRD data (see text). $\Theta_{\mathrm{E}}$ scatters around $\sim$ 60 K; the dashed line is a guide to the eye. (c) Debye temperature, $\Theta_{\mathrm{D}}$, versus $x$. $\Theta_{\mathrm{D}}$ increases slightly with increasing $x$. (d) The Sommerfeld coefficient, $\gamma$, obtained from fits using the Einstein model, versus $x$. The electronic specific heat coefficient $\gamma$ decreases as $x$ increases until $x =$ 3, and is enhanced as $x$ increases up to $x =$ 5. The value of $\gamma$ for PrPt$_{4}$Ge$_{12}$ in this work has a large error bar since we included the Einstein model; thus, the values obtained using other methods are also presented for reference~[\onlinecite{Gumeniuk08,Huang14}].}
\label{fig:Cp}
\end{figure}

Specific heat, $C$, versus $T$ data are shown in Fig.~\ref{fig:Cp}(a). The feature in $C/T$ associated with superconductivity was only observed in the $x =$ 0 and 0.5 samples (the $T_c$ values for $x \geq$ 1 are at temperatures below the low temperature limit of the PPMS DynaCool). Superconducting transition temperatures $T_c$ were determined with the aid of an idealized entropy-conserving constructions (not shown). The resultant $T_c$ values are consistent with the values obtained from the $\rho(T)$ and $\chi(T)$ measurements.
 
\indent In order to analyze the behavior of the electronic and phonon contributions to $C$, we first attempted to employ linear fits of $C/T$ versus $T^2$ to the data, which is a commonly-used method~[\onlinecite{Huang14}]. However, this method is inappropriate for PrPt$_4$Ge$_{12-x}$Sb$_x$; the electronic specific heat coefficient, $\gamma'$, becomes negative for $x \geq 4$, possibly due to low temperature upturns in the $C/T$ versus $T^2$ plots (data not shown). The Debye temperature, $\Theta'_{\mathrm{D}}$, extracted from the coefficient $\beta$ using the equation $C(T)/T = \gamma' + \beta T^2$, increases from $\sim$ 192 K to $\sim$ 271 K over the entire concentration range (data not shown). Possible explanations will be addressed below in the discussion section.

\indent Since we have evidence that the specific heat data need to be analyzed using an alternative method, we considered contributions due to rattling motion of Pr ions or a Pr nuclear Schottky. We reasoned that a nuclear Schottky contribution is not reasonable since it could not fully account for the continuous increase of the Debye temperature, $\Theta'_{\mathrm{D}}$, for all $x$. We found that introducing an Einstein contribution is more appropriate for the analysis in this study. For example, the Einstein model accurately describes the temperature dependence of the specific heat of filled Tl$_{0.22}$Co$_4$Sb$_{12}$ and unfilled Co$_4$Sb$_{12}$ skutterudite compounds~[\onlinecite{Sales99}]. Also, the broad kink observed in the $C/T$ versus $T^2$ data for NdOs$_4$Sb$_{12}$ was explained by using a combination of Debye and Einstein models~[\onlinecite{Ho05}], so there exists some precedent for including an Einstein model in analysis of specific heat data. If Pr ions behave as Einstein oscillators, the relationship between the thermal displacement parameter, $U_{\mathrm{iso}}$, and the Einstein temperature, $\Theta_{\mathrm{E}}$, is given by the expression: 
\begin{equation}
U_\mathrm{iso} = \frac{\hbar^2}{2m_{\mathrm{Pr}}k_{\mathrm{B}}\Theta_{\mathrm{E}}}\coth\left(\frac{\Theta_{\mathrm{E}}}{2T}\right),
\label{eq:Uiso}
\end{equation}
where $m_{\mathrm{Pr}}$ is the mass of the Pr ion and $k_{\mathrm{B}}$ is the Boltzman constant. For all $x$, estimated values of $\Theta_{\mathrm{E}}$ are $\sim$ 60 K, as shown in Fig.~\ref{fig:Cp}(b). Since there is support for the possibility that Pr ions are behaving like Einstein oscillators, the specific heat can be expressed as $C = \gamma T + C_{\mathrm{Ein}}(T) + C_{\mathrm{Deb}}(T)$~[\onlinecite{Ho05}], where
\begin{equation}
C_{\mathrm{Ein}}(T)= r\cdot 3R\frac{(\Theta_{\mathrm{E}}/T)^2 e^{(\Theta_{\mathrm{E}}/T)}}{(e^{(\Theta_{\mathrm{E}}/T)} -1)^2},
\label{eq:Cein}
\end{equation}
\begin{equation}
C_{\mathrm{Deb}}(T)= (17-r)\cdot \frac{12\pi^4}{5}R\left(\frac{T}{\Theta_{\mathrm{D}}}\right)^3,
\label{eq:Cdeb}
\end{equation}
$r$ is the mixing ratio of Pr ions, and $R$ is the universal gas constant. We assumed that only Pr ions are behaving like Einstein oscillators in this analysis; thus, the mixing ratio is constrained to be $r \leq$ 1. The least-squares fits of $\gamma T + C_{\mathrm{Ein}}(T) + C_{\mathrm{Deb}}(T)$ to the $C(T)$ data were performed in the temperature range between 1.8 and 30 K and corresponding best-fit values for $\gamma$, $\Theta_{\mathrm{E}}$, and $\Theta_{\mathrm{D}}$ were extracted. As shown in Fig.~\ref{fig:Cp}(b), $\Theta_{\mathrm{E}}$ scatters in the range of $\sim$ 60 to 65 K, indicating a consistent result with the values from the XRD analysis using Eq.~(\ref{eq:Uiso}). The $\Theta_{\mathrm{D}}$ versus $x$ plot in Fig.~\ref{fig:Cp}(c) exhibits a relatively modest increase of $\Theta_{\mathrm{D}}$ with $x$ compared to the analysis of $C(T)/T = \gamma' + \beta T^2$ fits to the specific data. The electronic specific heat coefficient $\gamma$, presented in Fig.~\ref{fig:Cp}(d), first decreases from $\sim$ 107 at $x =$ 0 to $\sim$ 1 mJ/mol K$^2$ at $x = 3$, and then increases to $\sim$ 180 mJ/mol K$^2$ for $x \geq 4$. Both $\Theta_{\mathrm{D}}$ and $\gamma$ values for $x =$ 0 are larger than previously reported values, $\gamma'$, of 48 and 87 mJ/mol K$^2$, respectively~[\onlinecite{Gumeniuk08,Huang14}]; however, this is expected due to the different methods used to extract the $\Theta_{\mathrm{D}}$ and $\gamma$ values.

\begin{figure}
\includegraphics[width=8.5cm]{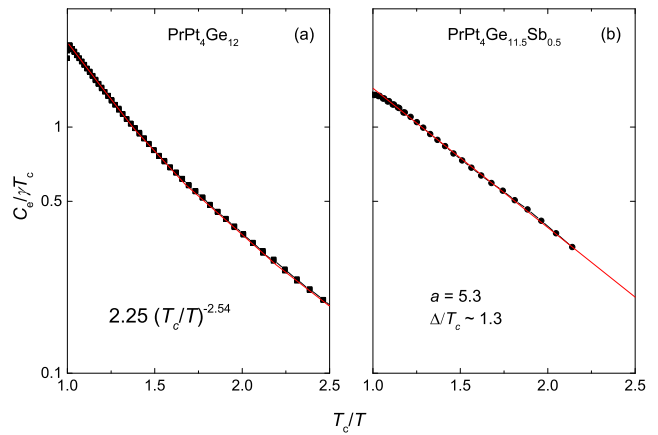}
\caption{(Color online) (a) and (b): Semilogarithmic plots of the electronic contribution to specific heat, $C_e / \gamma T_c$, in the superconducting state versus $T_c/T$ for PrPt$_4$Ge$_{12}$ and PrPt$_4$Ge$_{11.5}$Sb$_{0.5}$, respectively. The red lines represent the best fits to the data with 2.25$(T_c /T)^{-2.54}$ and $ae^{-\Delta/T_c}$, where $a =$ 5.3 and $\Delta/T_c =$ 1.3 for PrPt$_4$Ge$_{12}$ and PrPt$_4$Ge$_{11.5}$Sb$_{0.5}$, respectively.}
\label{fig:Ce}
\end{figure}

The electronic contribution to the specific heat, $C_{\mathrm{e}}(T)$, was extracted by subtracting the phonon contribution, $C_{\mathrm{ph}}(T) = C_{\mathrm{Ein}}(T) + C_{\mathrm{Deb}}(T)$, from the $C(T)$ data. Figure~\ref{fig:Ce} displays semilogarithmic plots of $C_e / \gamma T_c$ versus $T_c/T$ for PrPt$_4$Ge$_{12}$ and PrPt$_4$Ge$_{11.5}$Sb$_{0.5}$. The fits to the data were performed in the range 1 $\leq$ $T_c/T$ $\leq$ 2.5. The red line in Fig.~\ref{fig:Ce}(a) demonstrates that PrPt$_4$Ge$_{12}$ is well described by $b(T_c/T)^n$, where $n \sim -$2.5, suggesting multiband superconductivity or nodes in the gap function in this compound~[\onlinecite{Maisuradze09}]. The fit values for $b$ and $n$ are slightly different from the previous report~[\onlinecite{Huang14}], because different phonon contributions to the specific heat were subtracted. Figure~\ref{fig:Ce}(b) shows that the sample with $x =$ 0.5 can be fit by an exponential temperature dependence, $ae^{-\Delta/T_c}$, where $a$ is a fitting parameter and $\Delta$ is the superconducting energy gap. The coefficient $a$ is around 5.3 and $\Delta/T_c$ is $\sim$1.3, which is somewhat smaller than the value of 1.76 from the BCS prediction of weak-coupled superconductivity. This change from power law to exponential temperature dependence is similar to behaviors observed in previous reports for the Pr(Os$_{1-x}$Ru$_x$)$_4$Sb$_{12}$ and Pr$_{1-x}$Ce$_x$Pt$_4$Ge$_{12}$ systems~[\onlinecite{Frederick07, Huang14}] and could be explained by a crossover in the structure of the superconducting energy gap from one containing point-nodes to one that is nodeless, or a suppression of one or more superconducting energy gaps in a multiband superconductor~[\onlinecite{Maisuradze09, Sigrist91, Nakamura12, Zhang13}]. 

\section{Discussion}
\begin{figure}
\includegraphics[width=8cm]{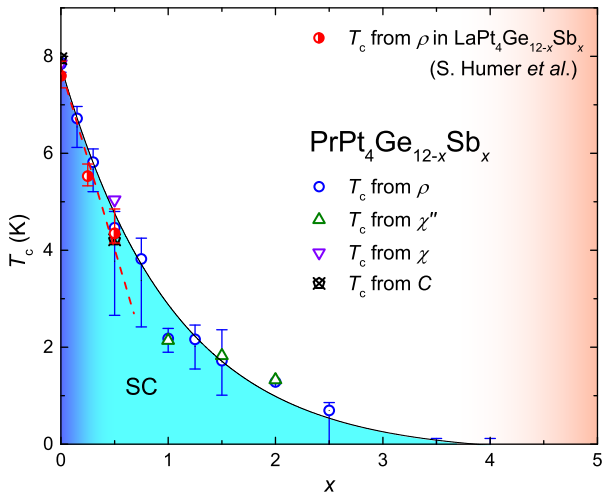}
\caption{(Color online) Phase diagram of the superconducting transition temperature $T_c$ as a function of Sb concentration $x$ based on electrical resistivity, AC and DC magnetic susceptibility, and specific heat measurements. The vertical bars for in the $T_c$ data represent the widths of the superconducting transitions and were derived from the $\rho(T)$ measurements as described in the text. $T_c$ is rapidly suppressed with $x$ up to $x \simeq$ 4, and only the onset of superconductivity is observed for $x \geq$ 3.5 in measurements performed down to $\sim$ 140 mK. The gradient-filled area of the SC region under the $T_c$ versus $x$ curve indicates the change of the superconducting energy gap. The red dashed line shows the suppression of superconductivity observed in LaPt$_4$Ge$_{12-x}$Sb$_x$~[\onlinecite{Humer13}].}
\label{fig:Tc}
\end{figure}

Figure~\ref{fig:Tc} summarizes results from $\rho(T)$, $\chi(T)$, $\chi''(T)$, and $C(T)$ measurements in a phase diagram of superconducting transition temperature $T_c$ versus nominal antimony concentration $x$. The $T_c$ values were taken from the onset of diamagnetic signals for the $\chi''(T)$ and $\chi(T)$ measurements. For the $C(T)$ measurements, $T_c$ was determined from the results of idealized entropy-conserving constructions~[\onlinecite{Bonjour91, Ota91}] (data not shown). This suppression of $T_c$ with $x$ is also consistent with the decrease of the upper critical field as shown in Fig.~\ref{fig:Hc2}(b).

\indent In order to understand the suppression of superconductivity, we need to consider the effects of Sb substitution in this system, including the increase of disorder, the increase in unit-cell volume, and the increase in electron concentration (electron doping). It has been reported that in Pt-Ge based skutterudite compounds, a large density of states (DOS) at the Fermi level is a common feature, and it is believed to facilitate superconductivity~[\onlinecite{Bauer07, Gumeniuk08}]. Moreover, soft x-ray valence-band photoemission spectroscopy on PrPt$_4$Ge$_{12}$ and LaPt$_4$Ge$_{12}$ display almost identical spectra, indicating that the electronic structures for both systems are very similar with one another~[\onlinecite{Nakamura10}]. Both have a similar $T_c$ near 8 K and exhibit evidence for multiband superconductivity; however, PrPt$_4$Ge$_{12}$ shows time-reversal symmetry breaking of the superconducting state~[\onlinecite{Maisuradze10}]. Therefore, a comparison of the suppression of superconductivity and the effect of Sb substitution on these two different systems, LaPt$_4$Ge$_{12}$ and PrPt$_4$Ge$_{12}$, may provide new insights. Our careful comparison reveals that there are several similarities and one significant difference~[\onlinecite{Humer13}]. The increase in disorder results in a rapid decrease of RRR (see Fig.~\ref{fig:RHO}), indicating a correlation between disorder and the suppression of superconductivity for low $x$ concentrations, consistent with the result in LaPt$_4$Ge$_{12-x}$Sb$_x$. The unit-cell volume of superconducting Pt-Ge based skutterudites seems to be uncorrelated or very weakly correlated to their superconducting states, because the changes of $T_c$ upon applied pressure are small or non-monotonic; $T_c$ is probably more dependent on the electronic structure~[\onlinecite{Khan08, Foroozani13}]. Humer \textit{et al}. discussed the decrease of the DOS and $\gamma$ values with Sb substitution, and its role in facilitating a rapid suppression of superconductivity~[\onlinecite{Humer13}]; a very similar decrease of $\gamma$ values for samples in the range 0 $\leq x \leq$ 3 is observed in this study . 

\indent The response of superconductivity to Sb substitution observed in this study contrasts with what was reported for LaPt$_4$Ge$_{12-x}$Sb$_x$. Even though there were no samples with Sb concentrations between $x =$ 0.5 and 3 in the LaPt$_4$Ge$_{12-x}$Sb$_x$ study, no evidence of superconductivity was observed at $x =$ 3 down to 0.4 K; moreover, a first principles calculation predicts that the system moves toward a metal-to insulator transition with higher Sb concentration. With these facts in mind, we conjecture that the suppression of superconductivity in LaPt$_4$Ge$_{12-x}$Sb$_x$ with $x$ could be faster than and/or different from that of PrPt$_4$Ge$_{12-x}$Sb$_x$, as indicated by the red dashed line in Fig.~\ref{fig:Tc}. Similarly, a different suppression rate of $T_c$ is observed upon substitution of Ce ions into the La and Pr sites~[\onlinecite{Huang15}]. The observed differences in the suppression of $T_c$, therefore, could be explained by differing pairing mechanisms in these two systems; PrPt$_4$Ge$_{12}$ exhibits an unconventional type of superconductivity, while LaPt$_4$Ge$_{12}$ is a conventional BCS-type superconductor. However, there is evidence that PrPt$_4$Ge$_{12}$ has multiple isotropic BCS-type energy gaps, which cannot be ruled out from this study~[\onlinecite{Chandra12, Nakamura12,Zhang13}]. Also, the expected behavior associated with point nodes in specific heat measurements on polycrystalline samples is not supported by the results from a single crystal study: Polycrystalline specimens show a $T^3$ temperature dependence of the electronic specific heat, which has been attributed to a possible nuclear Schottky anomaly arising from Pr-containing surface contamination~[\onlinecite{Zhang13}].

\indent A rattling mode of the rare-earth ion has been observed in several filled-skutterudite compounds~[\onlinecite{Keppens98,Nakai07}], especially in the $R$Os$_4$Sb$_{12}$ ($R=$ rare earth) family~[\onlinecite{Yanagisawa09,Yamaura11,Keiber12}]. On the other hand, the Pt-Ge based skutterudites exhibit no strong evidence for off-center displacements of filler ions; thus, the DOS, composed of Ge-$p$ and Pt-5$d$ states, has more significant effects on various phenomena~[\onlinecite{Gumeniuk10}]. However, the introduction of Gd ions into the La site in LaPt$_4$Ge$_{12}$ provides an extra phonon mode with $\Theta_{\mathrm{E}} \sim$ 24 K~[\onlinecite{Garcia12}]. In this work, a weak ``rattling'' mode with a value of $\Theta_{\mathrm{E}} \sim$ 60 K was estimated. In analogy with the behavior of Gd ions, the introduction of Sb ions into the Pt-Ge cage might produce additional phonon modes. The value of $\Theta_{\mathrm{E}}$ stays roughly constant throughout the entire range of $x$, suggesting that the rattling motion is independent of $x$; rattling in the PrPt$_4$Ge$_{12-x}$Sb$_x$ systems does not seem to have significant effects on either the superconducting state nor the localized Pr$^{+3}$ electronic configuration, consistent with the $x$-independent behavior of $\mu _{eff}/\mu _{B}$ as seen in Fig.~\ref{fig:mu_theta}(a).

\indent We are currently unable to definitively address the relationships among the Pr site occupancy, rattling dynamics, and the observed enhancement of electronic correlations for $x \geq$ 4 samples. It has been reported that the partial filling of the Pr sites does not have a significant effect on the superconducting properties of PrPt$_4$Ge$_{12}$~[\onlinecite{Venkateshwarlu14}]. A Pr Schottky anomaly could make it difficult to properly analyze the specific heat data~[\onlinecite{Zhang13}], possibly leading to an artificial enhancement of $\gamma$. The presence of small concentrations of impurities could also potentially be responsible for an anomalous enhancement of the specific heat at low temperature for higher $x$ (see Fig.~\ref{fig:Cp}(a)). All of these factors complicate our ability to clearly determine the phenomena or phase that occurs in the gradient-filled region for $x \geq$ 4 in Fig.~\ref{fig:Tc}. In order to have a better understanding of not only the features in the properties of samples with high Sb concentrations, but also the nature of superconductivity in PrPt$_4$Ge$_{12}$, future research such as neutron scattering, ultrasonic, or NMR-NQR measurements on high-quality single crystalline specimens will be necessary.  

\section{Concluding Remarks}
We have studied the superconducting and normal-state properties of PrPt$_{4}$Ge$_{12}$-based pseudoternary compounds in which Sb has been substituted for Ge. Polycrystalline samples of PrPt$_{4}$Ge$_{12-x}$Sb$_{x}$ with Sb concentrations up to $x =$ 5 were investigated via x-ray diffraction, electrical resistivity, magnetic susceptibility, and specific heat measurements. We observed a suppression of superconductivity with increasing Sb substitution up to $x =$ 4, above which, no signature of superconductivity was observed down to 140 mK. The electronic coefficient of specific heat $\gamma$ decreases with increasing Sb concentration in the superconducting region, indicating that the density of states might be an important parameter that facilitates superconductivity. The specific heat data for $x =$ 0.5 exhibits an exponential temperature dependence in the superconducting state, suggesting a nodeless superconducting energy gap. A constant ``rattling'' mode of Pr ions with a value of $\Theta_{\mathrm{E}} \sim$ 60 K across the entire substitution range was suggested by Rietveld refinements of XRD data and fits of the Einstein model to specific heat data; however, it does not seem to be correlated with superconductivity in this system.    

\begin{acknowledgments}
\indent This work was supported by the U. S. Department of Energy, Office of Basic Energy Sciences, Division of Materials Sciences and Engineering under Grant No. DE-FG02-04-ER46105 (characterization and physical properties measurements), and the National Science Foundation under Grant No. DMR 1206553 (low-temperature measurements). Helpful discussions with S. Ran are gratefully acknowledged.

\end{acknowledgments}

\bibliography{PrPt4Ge12-xSbx}

\end{document}